%% file: main.tex
\documentclass[11pt]{article}
\usepackage[utf8]{inputenc}
\usepackage[T1]{fontenc}
\usepackage{lmodern}
\usepackage[margin=2.5cm]{geometry}
\usepackage{amsmath}
\usepackage{graphicx}
\usepackage{booktabs}
\usepackage{array}
\usepackage{caption}
\usepackage{setspace}
\usepackage[authoryear,round]{natbib}
\usepackage{xcolor}
\usepackage[colorlinks=true,linkcolor=blue!50!black,citecolor=blue!50!black,urlcolor=blue!50!black]{hyperref}
\usepackage{url}
\usepackage{authblk} 
\newcommand{\doi}[1]{\href{https://doi.org/#1}{doi:\,\nolinkurl{#1}}} 

\graphicspath{{figures/}}
\title{Did the energy crisis turn borrowers green?\\Zonal power prices and green bond choice on Oslo B{\o}rs}

\author[1]{My Thi Diem Phan\thanks{Corresponding author. Email address diemmyphan@outlook.com.}} 
\author[2]{Tuyen Trung Truong} 
\author[3]{Hoai Phuong Ha} 
\author[4]{Dat Thanh Nguyen} 
\affil[1]{Independent researcher} 
\affil[2]{Department of Mathematics, University of Oslo} 
\affil[3]{Department of Computer Science, The Arctic University of Norway} 
\affil[4]{Faculty of Medicine, University of Oslo} 

\date{Working paper, October 2026}

\begin{document}
\maketitle

\begin{abstract}
\noindent The energy crisis of 2021 to 2023 caused a sharp increase in Nordic electricity prices, with substantial differences across bidding zones. We use this variation to examine whether issuers exposed to higher local electricity prices were more likely to choose green bonds over conventional bonds. We construct a dataset of bonds listed on Oslo B{\o}rs between 2019 and 2025 by combining ESMA FIRDS reference data, including matured bonds, with Euronext green bond classifications and issuer registry information. We then match issuers to their electricity bidding zones and examine the relationship between local power prices and green bond choice using issuer and quarter fixed effects. The share of green bonds among new listings increased from 15\% in 2019 to 48\% in 2024, but we find no statistically significant evidence that higher local electricity prices increased the likelihood of green bond issuance. This finding is consistent across price exposure windows of three to twelve months and for both utilities and other issuers. However, the estimates rely on 21 identifying issuers, and the minimum detectable effect at 80\% statistical power is 3.1 percentage points for a 10~EUR/MWh increase in electricity prices, meaning that smaller effects cannot be ruled out. The estimated effect is larger but imprecise among issuers with an earlier green bond, which implies an existing framework. We also find no clear association between carbon prices and green bond choice, although this analysis covers only eight utilities. Overall, the findings suggest that the growth of green bond issuance in Norway was more closely associated with broader market trends than with differences in local electricity prices.
\end{abstract}

\section{Introduction}

Green bonds accounted for an increasing share of new bond listings on Oslo B{\o}rs between 2019 and 2024, rising from 15\% to 48\%. This growth continued through the energy crisis of 2021 to 2023, when electricity prices reached record levels and the number of new bond listings fell by approximately one-third. These developments raise the question of whether higher energy costs encouraged firms to choose green bonds over conventional debt, or whether the growth in green issuance reflected broader changes in the bond market. We examine this question using a new dataset of Norwegian bond listings and variation in electricity prices across Nordic bidding zones.

Green bonds are often seen as a means of directing capital toward the low-carbon transition, and rising energy costs may strengthen the incentives for firms to invest in cleaner technologies. However, it is less clear whether these costs also influence the type of debt firms issue. The Nordic electricity market provides a useful setting for examining this relationship because Norway has five electricity bidding zones (NO1 to NO5), while Sweden has four, each with its own market price. During the energy crisis, prices increased sharply in southern zones but remained substantially lower in northern zones. Firms issuing bonds in the same quarter could therefore face different local electricity prices, allowing us to examine whether this variation is associated with their choice between green and conventional bonds.

To investigate this relationship, we construct a dataset that includes both active and matured bonds listed on Oslo B{\o}rs between 2019 and 2025. Existing exchange lists generally report currently listed instruments, which can exclude bonds issued in earlier years that have since matured. Relying on these lists alone may therefore introduce survivorship bias and distort the observed growth of green issuance \citep{brown1992}. We address this concern by combining the European Securities and Markets Authority's Financial Instruments Reference Data System (FIRDS), which retains records of terminated instruments, with Euronext green bond classifications. We also verify green labels against issuer reports and obtain issuer characteristics and locations from the GLEIF and Br{\o}nn{\o}ysund registers. The resulting dataset contains 1,719 bonds issued by 334 issuers. It shows that green bonds represented 36\% of new listings in both 2021 and 2022, even though electricity prices reached approximately 167~EUR/MWh in the period preceding listings in 2022 and the total number of new listings declined by one-third.

Our empirical analysis uses issuer and quarter fixed effects to examine whether changes in local electricity prices are associated with green bond choice. Issuer fixed effects account for time-invariant differences between firms, while quarter fixed effects absorb market-wide developments, including changes in interest rates and the common effects of the energy crisis. The identification therefore relies on differences in local electricity prices across bidding zones within the same quarter. We find a small and statistically insignificant relationship, with an estimated coefficient of 0.002 per 10~EUR/MWh and a standard error of 0.011. A placebo test that randomly reassigns bidding zones places the observed estimate at the 26th percentile of the placebo distribution. Although a specification with year fixed effects produces a small positive association, the placebo results suggest that this association may reflect common time trends rather than differences in local electricity prices. The findings are also consistent across alternative price exposure windows of six and twelve months. We find no clear relationship between carbon prices and green bond choice, including among utilities, although the carbon price analysis is based on only eight utilities.

These results should be interpreted in light of the limited variation available for identification. Only 21 issuers have both green and conventional bonds and an assigned electricity price zone, which leads to considerable uncertainty around the estimates. For a price increase comparable to the 2022 shock, the 95\% confidence interval ranges from $-36$ to $+47$ percentage points, while the minimum detectable effect at 80\% statistical power is approximately 40 percentage points. We therefore cannot rule out economically meaningful effects, even though the estimated average relationship is close to zero. Among issuers with an existing green bond framework, the estimated effect is larger but remains statistically insignificant, and this analysis relies on only eight issuers.

This study contributes to the literature in three ways. First, we provide a historical dataset of bond listings on Oslo B{\o}rs that includes matured instruments and reduces the risk of survivorship bias. This is particularly important when examining changes in green bond issuance over time, as datasets limited to currently listed bonds may misrepresent earlier issuance activity \citep{brown1992}. Second, we extend the literature on green bonds by examining the choice of financing instrument rather than its pricing or investor demand. Previous research has studied the green bond premium \citep{zerbib2019,larcker2020}, investor ownership and preferences \citep{baker2022,sangiorgi2021}, and the relationship between green bond issuance, stock market reactions, and issuer environmental performance \citep{flammer2021,tang2020,fatica2021bse}. Nordic studies have also highlighted the importance of institutional relationships, business culture, and issuer motivations in green bond markets \citep{maltais2020,torvanger2021}. We add to this literature by examining whether local energy price exposure is associated with the decision to issue green rather than conventional debt.

Third, our analysis connects green bond issuance to research on energy prices and firms' investment decisions. Previous studies show that emission costs are reflected in wholesale electricity prices \citep{fabra2014}, while energy prices can influence innovation and investment as firms adjust their capital stock \citep{popp2002,linn2008,steinbuks2014}. Although these mechanisms suggest that higher energy costs may encourage investment in cleaner technologies, they do not necessarily imply an immediate change in financing choices. Our findings provide no clear evidence of such a response in green bond issuance within a year, but the relatively small number of identifying issuers limits the precision of this conclusion. To account for the small number of clusters, we use wild cluster bootstrap inference following \citet{mackinnon2023,webb2023}.

\section{Data}

\subsection{The Norwegian bond market}

Oslo B{\o}rs, now part of Euronext, operates two main bond markets: the regulated Euronext Oslo B{\o}rs (XOSL) and the unregulated Euronext ABM, formerly Nordic ABM \citep{euronext_oslo,euronext_abm}. The regulated market follows European Economic Area (EEA) securities rules, and trading venues report instrument reference data to the European Securities and Markets Authority (ESMA) through the Financial Instruments Reference Data System (FIRDS) \citep{euronext_oslo,esma_mifir27}. In contrast, Euronext ABM operates under rules set by Oslo B{\o}rs and has lighter listing requirements, including no requirement for a full EEA prospectus \citep{euronext_abm}.


The Norwegian bond market relies on a common trustee, and Nordic Trustee reported in 2015 that it acted as bond trustee for about 95\% of Norwegian bond issues, representing the bondholders under standard bond terms \citep{nordictrustee2015}. Utilities are the largest sector in the Norwegian investment grade corporate bond market \citep{nordictrustee2026}, and most hydropower in Norway is owned by the state, municipalities and counties \citep{nve_ownership}. The assets of power companies, such as hydropower plants and grids, have long lives, and this may explain why they borrow so much in the bond market.

Green bonds are generally issued under a framework based on the four core components of the International Capital Market Association's Green Bond Principles, including the process used to identify and select eligible projects \citep{icma2025gbp}. Although these principles are voluntary, they recommend an external review, commonly provided through a second-party opinion \citep{icma2025gbp,icma2020er}. CICERO Shades of Green, an Oslo-based provider acquired by S\&P Global in December 2022, has been an important provider of such reviews \citep{spglobal2022}. Once a framework is established, an issuer can use it for subsequent green bond offerings \citep{cicero2020glitre}.

The validity of these reviews varies across frameworks. Among six CICERO opinions issued between 2016 and 2020 that we examined, the most recent specifies a three-year validity period if the framework remains unchanged \citep{cicero2020glitre}. Two others require reassessment when the framework changes, while three do not specify a validity period \citep{cicero2017eidsiva,cicero2019sbab,cicero2016kbn,cicero2016entra,cicero2017vasakronan}. Since the acquisition, S\&P Global has described its opinions as assessments made at a particular point in time, without subsequent monitoring \citep{spglobal2022,sp2025ferde}.

\subsection{Bond universe}

We construct a dataset of bonds listed on the regulated Oslo B{\o}rs market (MIC code XOSL) with a first trading day between January 2019 and December 2025. The dataset includes bonds that have since matured, allowing us to examine historical issuance without restricting the sample to instruments that remain listed. We combine five public data sources, summarised in Table~\ref{tab:sources}.

FIRDS contains 3,520 debt instrument records on XOSL, of which 1,969 are classified as terminated. These records are important because bonds listed in earlier years remain observable after redemption. Restricting the data to bonds with a first trading day between 2019 and 2025 gives 1,719 bonds from 334 issuers.

FIRDS has two limitations that matter for our analysis. First, it reports the first trading day rather than the issue date, and listing may occur several weeks after settlement. Among the 286 ESG bonds matched to Euronext listing records, the two dates coincide for 87\% of observations. Second, FIRDS reports the latest total nominal amount, which may include subsequent tap issues rather than only the original issue amount. We therefore use the reported nominal values with this limitation in mind. Three records contain amounts approximately one million times larger than expected across trading venues. Instead of applying an assumed correction, we treat nominal amounts above EUR 20 billion as missing. All monetary amounts are converted into euros using the European Central Bank's daily exchange rates.

\begin{table}[htbp]
\centering
\small
\caption{Data sources. All sources were accessed in October 2026.}
\label{tab:sources}
\begin{tabular}{>{\raggedright\arraybackslash}p{3.4cm}>{\raggedright\arraybackslash}p{6.6cm}>{\raggedright\arraybackslash}p{4.2cm}}
\toprule
Source & What it gives & Used for \\
\midrule
\href{https://registers.esma.europa.eu/publication/searchRegister?core=esma_registers_firds}{ESMA FIRDS} & All debt instruments on XOSL, with first trading day, maturity, total nominal, termination date and issuer LEI. Terminated bonds are kept. & Bond universe (green and conventional) \\
\href{https://live.euronext.com/en/products/fixed-income/esg-bonds}{Euronext ESG bond list} & ISIN, issuer and bond type (green, social, sustainability, sustainability linked) & Green label \\
\href{https://www.gleif.org}{GLEIF} and \href{https://www.brreg.no}{Br{\o}nn{\o}ysund Register Centre} & Organisation number, NACE code, legal form and registered address & Issuer type and headquarters location \\
\href{https://transparency.entsoe.eu}{ENTSO-E Transparency Platform} & Hourly prices in the day-ahead market for NO1 to NO5, SE1 to SE4, DK1, DK2, FI, and Germany with Luxembourg & Zonal power price \\
\href{https://ember-energy.org}{Ember}, \href{https://www.norges-bank.no}{Norges Bank} and \href{https://data.ecb.europa.eu}{ECB} & Daily EU ETS price (front December futures, based on Montel data), policy rate and exchange rates & Carbon price, rate control and conversion to EUR \\
\bottomrule
\end{tabular}
\end{table}

We restrict the bond universe to XOSL because Euronext's other Oslo trading venues, Expand and Growth, have no bond records in FIRDS. Some ESG bonds listed in Oslo are also absent from FIRDS, possibly because they are listed on Nordic ABM, which does not report to this system. These instruments are excluded from both green and conventional bond groups so that all observations come from the same trading venue.

\subsection{Green label}

We identify green bonds by matching their International Securities Identification Numbers (ISINs) to instruments classified as ``Green Bond'' on Euronext's ESG bond list. This external classification is necessary because the instrument names reported in FIRDS do not provide a reliable indication of green status.

We assess the completeness and accuracy of the classification in two ways. First, 53 of the 225 matched green bonds are recorded as terminated in FIRDS, indicating that the Euronext list includes at least some matured green bonds. Second, we compare the Euronext classifications with issuer green bond reports for Hafslund, Statnett, Lyse, Entra, and Sparebanken M{\o}re. Of the 59 bonds examined, 54 can be matched by bond number, while the remaining five are confirmed through issuer press releases. We find no incorrectly labelled bonds in this validation sample. However, two Entra green bonds issued in July 2026, outside our study period, are absent from the Euronext list, suggesting that the classification may not be exhaustive.

\subsection{Issuers and price zones}

We identify issuers using their Legal Entity Identifiers (LEIs) and match these records to the Global Legal Entity Identifier Foundation (GLEIF) database. For Norwegian issuers, we also use organisation numbers to obtain information from the Br{\o}nn{\o}ysund Register Centre. Issuers are classified as utilities, banks, municipalities, real estate firms, or other entities based on industry codes, legal form, and sector information. We apply 18 documented adjustments at the issuer level and two at the bond level to correct classifications where necessary. Utilities are further divided into electricity producers, grid companies, and integrated firms. For issuers affected by mergers, we retain the entity responsible for issuing the bond at the time of issuance rather than its current successor.

We assign electricity bidding zones using issuers' registered headquarters. Norwegian issuers are matched by postcode to zones NO1 to NO5, while Swedish and Danish issuers are assigned to their respective zones using headquarters regions. Issuers in Finland and Germany are assigned to the corresponding national price areas, with Germany including Luxembourg. We also flag headquarters located near bidding zone boundaries. For national firms headquartered in Oslo, the local bidding zone may not accurately reflect the geographic distribution of their operations or electricity consumption. We discuss this measurement limitation in Section~\ref{sec:limits}.

Our main explanatory variable is the average hourly day-ahead electricity price in the issuer's assigned bidding zone over the three months preceding the bond's first trading day. We use a similar three-month average for carbon prices, measured using Ember's daily European Union Emissions Trading System (EU ETS) front-December futures series in EUR per tonne. Between 2020 and 2025, annual average prices in this series differ from European Energy Exchange (EEX) primary auction prices by approximately 1.5\% to 2.5\%. Comparable EEX files for 2018 and 2019 were unavailable, and confirmation of the licence for the Ember series remains pending. The monetary policy control is the Norges Bank policy rate, averaged over the same three-month period.

\subsection{Main sample}

We exclude municipalities, banks, and money market certificates from the main estimation sample because their financing arrangements differ from those of other corporate issuers. Municipalities account for 856 conventional bonds and six green bonds on XOSL, with borrowing decisions often linked to public budget cycles. Banks are excluded because many of their green covered bonds are listed outside XOSL, while money market certificates mainly serve short-term refinancing needs. The latter group contains 63 conventional instruments and no green bonds.

These exclusions leave 599 bonds, of which 208 are green. After restricting the sample to green and conventional bonds with an assigned electricity bidding zone, the estimation sample contains 513 bonds from 136 issuers, including 202 green bonds. Of these issuers, 21 have issued both green and conventional bonds and therefore contribute to the within-issuer identification of green bond choice. We retain the excluded groups for robustness analyses, while Table~\ref{tab:sample} summarises the sample construction.

\begin{table}[htbp]
\centering
\small
\caption{Sample construction.}
\label{tab:sample}
\begin{tabular}{lrr}
\toprule
Step & Bonds & Green \\
\midrule
XOSL bonds with first trading day in 2019 to 2025 & 1,719 & 225 \\
Main sample (without municipalities, banks and certificates) & 599 & 208 \\
Estimation sample (green or conventional, with a price zone) & 513 & 202 \\
Issuers in the estimation sample, and issuers with both types & 136 and 21 & \\
\bottomrule
\end{tabular}
\end{table}

\section{Descriptive facts}

Green bonds became a much larger share of new listings on Oslo B{\o}rs between 2019 and 2024, rising from 15\% to 48\%. The main sample contains 599 bonds from 187 issuers, with a total reported volume of EUR 55.9 billion, including 208 green bonds. Green bonds account for 36.0\% of green and conventional issues by number and 29.2\% by volume. Table~\ref{tab:descriptive} describes the sample by listing year, issuer type, and headquarters country.

\begin{figure}[htbp]
\centering
\includegraphics[width=0.8\textwidth]{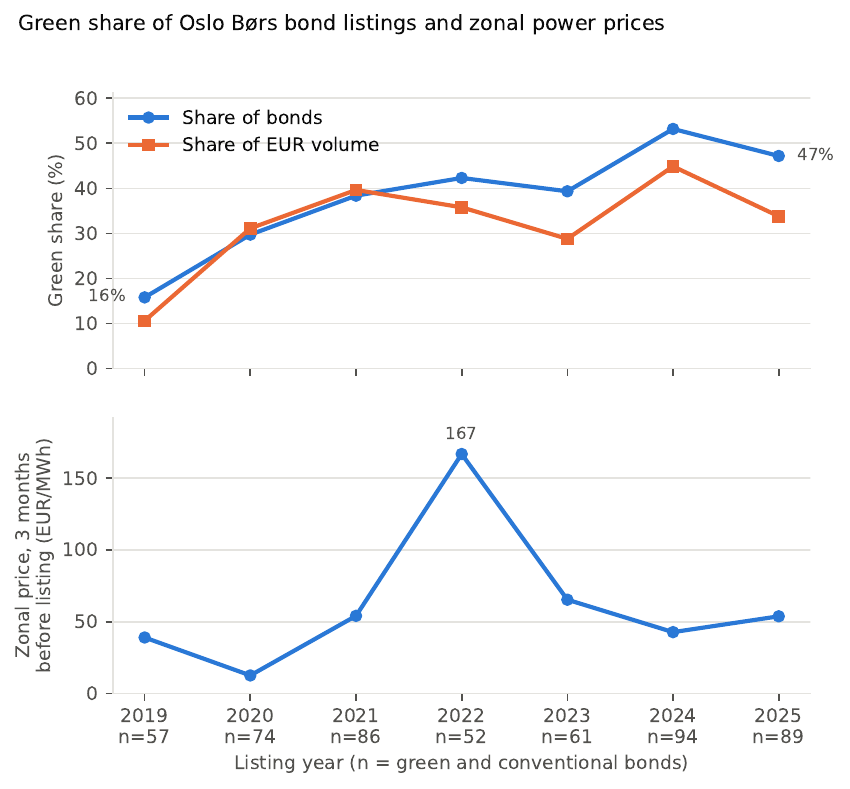}
\caption{Green share of bond listings on Oslo B{\o}rs (top) and mean zonal day ahead price in the three months before listing (bottom), by listing year. The figure covers bonds in the main sample whose issuers have a price zone, and $n$ is the number of green and conventional bonds per year. Shares differ slightly from Table 3, which also includes issuers without a price zone.} 
\label{fig:share}
\end{figure}

The first pattern is that bond issuance declined during the energy crisis without a comparable fall in the green share. New listings fell from 103 in 2021 to 70 in 2022, a decline of 32\%, while reported volume decreased from EUR 9.9 billion to EUR 7.3 billion. Green listings declined from 34 to 24 and conventional listings from 61 to 42, leaving the green share among these two types almost unchanged at 35.8\% and 36.4\%, respectively. Meanwhile, the average zonal electricity price over the three months preceding listing reached approximately 167~EUR/MWh in 2022 among bonds for which a zone could be assigned (Figure~\ref{fig:share}).

\begin{table}[htbp]
\centering
\caption{Main sample by listing year, issuer type and country of the headquarters. Bonds and volume include other ESG bonds, but both green shares use green plus conventional bonds as the denominator, as in the models.} 
\label{tab:descriptive}
\resizebox{\textwidth}{!}{\input{tables/table1.tex}}
\end{table}

The second pattern concerns differences across issuers. Utilities issued 95 green bonds among 127 green and conventional bonds (74.8\%), compared with green shares of 37.0\% for real estate firms and 21.4\% for other issuers. This is a notable contrast with interviews conducted between 2013 and 2019, when participants in the Norwegian energy sector often viewed green bonds as unnecessary for assets that were already low in carbon \citep{torvanger2021}. In the more recent sample, green bonds had become a common financing choice for utilities, although these descriptive figures alone do not explain why.

Issuance patterns also differ by the issuer's headquarters country. Swedish issuers account for 27 green bonds among 37 green and conventional listings (73.0\%), largely concentrated in one property company, whereas none of the 13 Danish bonds is green. These observations represent firms choosing to list in Oslo rather than the wider Swedish or Danish bond markets. Figure~\ref{fig:share} compares annual green shares with average zonal prices before listing. Although the peak in electricity prices in 2022 is not accompanied by a comparable change in the green share, annual comparisons cannot isolate local price exposure from broader market developments. The empirical design in Section~\ref{sec:design} addresses this distinction.

\section{Design}
\label{sec:design}

Our main analysis examines whether a bond is more likely to be green when its issuer has recently faced higher electricity prices in its assigned bidding zone. The specifications were recorded before estimation. We observe bond $i$ issued by firm $j$ and listed in quarter $t$, and define the dependent variable as one for a green bond and zero for a conventional bond. Other environmental, social, and governance (ESG) bond types are excluded. The main explanatory variable is the average day-ahead price in the issuer's assigned zone during the three months before listing, measured in units of 10~EUR/MWh. We estimate the following linear probability model:
\begin{equation}
\text{Green}_{ijt} = \beta \, P_{z(j),t} + \alpha_j + \gamma_t + \varepsilon_{ijt}.
\end{equation}

We consider two specifications that differ in their treatment of common time variation. Specification A includes issuer and year fixed effects and controls for the Norges Bank policy rate, leaving within-year price variation that may also reflect broader shocks. Our preferred specification, Specification B, includes issuer and quarter fixed effects, absorbing factors common to issuers in a given quarter. Its coefficient is therefore estimated only from differences in zonal prices between issuers in the same quarter, rather than from the overall rise and fall in electricity prices during the crisis. Within-issuer changes in the bond label are central to identification, and 21 issuers in the estimation sample have issued both green and conventional bonds. For the carbon-price analysis, we add the mean price of European Union emissions allowances over the preceding three months and interact it with an indicator for utilities, retaining issuer and quarter fixed effects.

Standard errors are clustered at the issuer level. Given the small number of identifying issuers, we report wild cluster bootstrap $p$ values using Webb weights and 9,999 draws, following \citet{mackinnon2023} and \citet{webb2023}, and obtain confidence intervals by inverting the bootstrap test. We also conduct 1,000 placebo reassignments of electricity zones among Norwegian issuers to assess whether the observed estimates are unusual relative to alternative zone assignments. To convey the precision of the main estimate, we calculate the minimum detectable effect at 80\% power for a two-sided 5\% test, approximated as 2.8 times the standard error. We additionally express the estimates for a 130.7~EUR/MWh shock, the difference between average pre-listing prices in 2022 and in 2019 to 2021.

Robustness checks include conditional logit, exclusion of the largest Swedish issuer (Vasakronan), restriction to Norwegian issuers, and inclusion of municipalities and banks. We also examine the number and volume of green and conventional bonds in a zone-quarter panel, using Poisson models for counts and Poisson pseudo-maximum likelihood (PPML) for euro-denominated volumes, with zone and quarter fixed effects.

\subsection{Threats to identification}

The preferred specification reduces the influence of common market developments, but several sources of bias remain possible. First, developments affecting one zone within a quarter may influence both electricity prices and firms' financing choices. For example, a regional expansion could raise electricity demand while also changing firms' borrowing needs. The placebo exercise helps assess whether the estimated zonal association is distinctive, although it cannot eliminate such confounding factors.

Second, our price measure may contains error, because the zone of the headquarters measures exposure only roughly for firms that operate in several zones, firms may look at forward prices rather than recent spot prices, and the decision to borrow may come before the first trading day recorded in FIRDS. If this error is unrelated to the choice of bond type, it biases the estimate toward zero, so part of the null result may reflect attenuation. Error that varies with the choice of bond type over time could bias the estimate in either direction, although the issuer fixed effects absorb any part of it that is constant for a given issuer.
Third, reverse causality is unlikely, because issuers do not choose the zone of their headquarters in order to issue green bonds, and a single bond issue cannot move the zonal price.
This does not make the estimates causal, and we interpret them as conditional associations.

\section{Results}

Table~\ref{tab:results} reports the main estimates as changes in the probability of green bond issuance for a 10~EUR/MWh increase in zonal electricity prices. In Specification A, the estimated coefficient is 0.0173, corresponding to 1.73 percentage points, with a bootstrap $p$ value of 0.096. The positive association is therefore not statistically significant at the 5\% level, and the placebo analysis below indicates that it is not distinctive of the actual zonal assignments.

\begin{table}[htbp]
\centering
\small
\caption{Linear probability models with outcome 1 if the bond is green. The sample has 513 bonds and 136 issuers, of which 21 identify the effect (8 utilities in the carbon model). Standard errors are clustered by issuer, and p values and confidence intervals come from a wild cluster bootstrap with Webb weights and 9,999 draws.}
\label{tab:results}
\resizebox{\textwidth}{!}{%
\begin{tabular}{llrrrc}
\toprule
Model & Term & Coef. & SE & Bootstrap $p$ & 95\% CI \\
\midrule
Spec A with issuer and year FE & Zonal price & 0.0173 & 0.0086 & 0.096 & [$-0.0041$, 0.0319] \\
Spec A & Policy rate (pp) & $-0.0334$ & 0.0895 & 0.687 & [$-0.2050$, 0.1511] \\
\textbf{Spec B with issuer and quarter FE} & \textbf{Zonal price} & \textbf{0.0024} & \textbf{0.0111} & \textbf{0.809} & \textbf{[$\mathbf{-0.0277}$, 0.0361]} \\
Carbon with issuer and quarter FE & EUA (per 10 EUR/t) & $-0.0857$ & 0.0960 & 0.324 & [$-0.2712$, 0.0810] \\
Carbon & EUA $\times$ utility & 0.0580 & 0.0373 & 0.177 & [$-0.0257$, 0.1400] \\
\bottomrule
\end{tabular}}
\end{table}

Once quarter fixed effects are included in Specification B, the estimate falls to 0.0024, with a bootstrap $p$ value of 0.809. This corresponds to an estimated 0.24 percentage point difference for a 10~EUR/MWh higher price. Scaling the coefficient to the 130.7~EUR/MWh difference between 2022 and the earlier period gives an estimated 3.2 percentage point effect, but the 95\% confidence interval ranges from $-36$ to $+47$ percentage points. The minimum detectable effect at 80\% power is approximately 3.1 percentage points per 10~EUR/MWh, or around 40 percentage points for a shock of this size. These intervals leave substantial uncertainty about economically meaningful effects.

We also find no statistically significant relationship between emissions allowance prices and green bond choice ($p = 0.324$). The estimated interaction between the allowance price and the utility indicator is positive, but imprecise ($p = 0.177$), and only eight utilities provide within-issuer variation in bond type for this analysis. Taken together, the estimates do not establish a link between recent local electricity prices and green bond choice, while the limited precision prevents us from concluding that such a link is absent.

\section{Robustness and placebo}
\label{sec:robust}

The preferred specification remains close to zero across the main robustness checks in Table~\ref{tab:robust}. The placebo exercises also help explain the difference between Specifications A and B. When electricity zones are randomly reassigned, Specification A produces an average coefficient of 0.0167, close to its observed estimate of 0.0173, which lies at the 51st percentile of the placebo distribution. This suggests that much of the association in Specification A is related to price movements shared across zones rather than to their actual assignment to issuers. For Specification B, the observed estimate lies at the 26th percentile, providing no indication that the true zone assignments produce an unusually large coefficient.

The conditional logit model produces a positive coefficient for Specification A (0.170 in log odds, $p = 0.047$), based on analytical rather than bootstrap inference. This result is less persuasive given the placebo findings for that specification. A corresponding conditional logit for Specification B could not be estimated because the 26 quarter indicators perfectly separate outcomes among the identifying issuers. We report this limitation instead of selectively dropping quarter indicators.

\begin{table}[htbp]
\centering
\small
\caption{Robustness checks, all fixed before estimation. The p values come from the wild cluster bootstrap unless noted otherwise. For R5, $p$ is the share of 1,000 placebo draws that are at least as large in absolute value as the true estimate. R6 uses analytical standard errors clustered by issuer and by quarter \citep{cameron2011,petersen2009}, with p values from a t distribution with 27 degrees of freedom, and not the bootstrap.} 
\label{tab:robust}
\resizebox{\textwidth}{!}{%
\begin{tabular}{lrrrrr}
\toprule
Check & Spec A coef. & Spec A $p$ & Spec B coef. & Spec B $p$ & Bonds and identifying issuers \\
\midrule
R1 Conditional logit (log odds) & 0.170 & 0.047 (Wald) & not estimable & & 216 and 21 \\
R2 Without Vasakronan & 0.0165 & 0.127 & 0.0006 & 0.954 & 493 and 20 \\
R3 Norwegian issuers only & 0.0173 & 0.119 & $-0.0001$ & 0.995 & 456 and 19 \\
R4 With municipalities and banks & 0.0024 & 0.281 & $-0.0002$ & 0.952 & 1,570 and 29 \\
R5 Zone placebo, percentile of the true estimate & 51.1 & 0.489 & 26.1 & 0.847 & 513 and 21 \\
R6 Clustered by issuer and quarter & 0.0173 & 0.063 & 0.0024 & 0.829 & 513 and 21 \\ 
\bottomrule
\end{tabular}}
\end{table}

The secondary outcomes tell a similar, although less precise, story. In the zone-quarter panel, a 10~EUR/MWh increase in lagged prices is associated with 1.4\% fewer green bonds ($p = 0.71$) and 2.0\% lower green bond volume ($p = 0.56$). Green issuance occurs in only six zones, making these estimates difficult to interpret precisely. Conventional bond counts decline by an estimated 4.9\%, with $p = 0.067$ when standard errors are clustered by zone, compared with $p = 0.049$ under heteroskedasticity-robust errors. Given the small number of zones and the eight secondary tests considered, we treat this result as suggestive rather than robust. Conventional bond volume shows no significant association ($p = 0.65$).

Higher electricity prices could affect utilities and other issuers differently. Utilities, which sell power, may receive additional revenue when prices increase, which could reduce their need for external finance, whereas electricity-consuming firms may face higher costs and stronger incentives to invest in efficiency. These responses could offset one another in the pooled estimate. We therefore interact zonal prices with an indicator for utilities in Specification B. As shown in Table~\ref{tab:mechanism}, the coefficients are close to zero for other issuers ($-0.0026$) and utilities (0.0031), and their difference is not statistically significant. Although the decline in conventional bond counts is compatible with reduced borrowing after higher utility revenues, the count model pools issuer types and contains only ten zones, so it cannot establish this mechanism.

\begin{table}[htbp] 
\centering 
\small 
\caption{Effect of the zonal price for sellers and buyers of power, from Spec B with the zonal price interacted with an indicator for utilities. Coefficients are per 10~EUR/MWh, and the sample has 513 bonds and 136 issuers. Standard errors are clustered by issuer, and p values and confidence intervals come from a wild cluster bootstrap with Webb weights and 9,999 draws.} 
\label{tab:mechanism} 
\resizebox{\textwidth}{!}{
\begin{tabular}{lrrrcr} 
\toprule 
Group & Coef. & SE & Bootstrap $p$ & 95\% CI & Identifying issuers \\ 
\midrule 
Other issuers (buyers of power) & $-0.0026$ & 0.0160 & 0.856 & [$-0.0341$, 0.0314] & 13 \\ 
Utilities (sellers of power) & 0.0031 & 0.0107 & 0.742 & [$-0.0254$, 0.0417] & 8 \\ 
Difference between utilities and other issuers & 0.0057 & 0.0130 & 0.684 & [$-0.0214$, 0.0429] & 21 \\ 
\bottomrule 
\end{tabular}} 
\end{table}

\section{Is it a matter of time?}
\label{sec:timing}

One possible explanation for the weak short-run association is that firms take time to adjust their investment and financing decisions after an energy price shock. Previous research shows that energy prices can influence innovation \citep{popp2002}, technology choices in new plants \citep{linn2008}, and the efficiency of capital as older equipment is replaced \citep{steinbuks2014}. When future prices are uncertain, the option to wait can further delay irreversible investment \citep{dixit1994,bloom2009}. These mechanisms imply a gradual investment response, but the financing decision need not follow the same timing because green bonds can also refinance existing eligible assets, including hydropower facilities.

Our baseline exposure measure averages electricity prices over the three months preceding the first trading day, which may be later than the date when financing was arranged. We searched public notices for issue dates for 20 bonds and located ten; in these cases, issuance preceded listing by a median of approximately 65 days. To assess whether the choice of price window matters, Table~\ref{tab:timing} reports six- and twelve-month averages and a window covering months five to two before listing. The estimated coefficients in Specification B remain close to zero across these alternatives. Specification A produces coefficients of similar magnitude, but their precision decreases as the window lengthens, partly because twelve-month average prices vary relatively little within a calendar year.

\begin{table}[htbp]
\centering
\small
\caption{Zonal price windows before listing. Coefficients per 10~EUR/MWh with wild cluster bootstrap $p$ values in parentheses. Every model has 513 bonds, 136 issuers and 21 identifying issuers.}
\label{tab:timing}
\begin{tabular}{lcc}
\toprule
Price window before listing & Spec A & Spec B \\
\midrule
3 months & 0.017 (0.096) & 0.002 (0.80) \\
6 months & 0.016 (0.17) & $-0.002$ (0.89) \\
12 months & 0.016 (0.32) & $-0.011$ (0.50) \\
Months 5 to 2 & 0.011 (0.43) & $-0.020$ (0.49) \\
\bottomrule
\end{tabular}
\end{table}

A first green bond usually requires an issuer to prepare a green bond framework and obtain an external review, so an immediate response may not be feasible for firms without such arrangements. However, 17 of the 21 identifying issuers had previously listed a green bond before at least one later issue, covering 164 of their 216 bonds. Restricting the estimation to bonds issued after an issuer's first green bond gives a larger Specification B coefficient of 0.033, although it remains statistically insignificant at the 5\% level (bootstrap $p = 0.089$, 95\% confidence interval from $-0.037$ to 0.101). Moreover, only eight issuers identify this restricted estimate.

We find no clear evidence of a delayed increase in first green bond issuance in high-price zones. Among Norwegian issuers, first green listings in these zones number 16 before 2022 and 15 in 2022 or later, compared with three and two in low-price zones. The green share in high-price zones rises from 23.9\% in 2019 to 2021 to 49.3\% in 2023 to 2025, but low-price zones contain only 14 to 15 bonds per period, too few for a reliable comparison.

These findings do not provide strong support for an adjustment delay within the observed period, but several explanations remain possible. The financing response may take longer than the sample allows us to observe, or green bond issuance may be shaped by established frameworks and market practices rather than by recent local costs. Investor demand is another possibility, as funds with green mandates may purchase eligible bonds irrespective of an issuer's immediate electricity costs. This interpretation is consistent with evidence on green bond ownership and institutional investor preferences \citep{baker2022,sangiorgi2021}. Our data do not identify bond purchasers or primary-market pricing, so we cannot distinguish these explanations without allocation or investor holdings data. The possible role of forward electricity prices is discussed in Section~\ref{sec:limits}.

\section{Limitations}
\label{sec:limits}

The main limitation is statistical power, because although the estimation sample contains 513 bonds, only 21 issuers have both green and conventional bonds and an assigned price zone, and most of them are in NO1. 
Several other zones, including NO4, SE4, DK1, DK2 and FI, contribute very few green bonds. 
For a shock comparable to the price increase of 2022, the minimum detectable effect at 80\% power is about 40 percentage points, and the 95\% confidence interval runs from $-36$ to $+47$ percentage points. 
The results therefore cannot exclude effects that would be economically important. 
An issuer level model of the timing of the first green bond could offer complementary evidence, but there are too few first green bonds across zones for such a model to be informative. 

A related limitation is that the price differences between zones within a quarter, which identify Spec B, are much smaller than the overall price shock of 2022. 
Scaling the Spec B coefficient to an increase of 130.7~EUR/MWh therefore assumes a linear relationship well beyond the variation that identifies it. 
Bonds on Nasdaq Stockholm, Copenhagen and Helsinki are reported to FIRDS and could be added from public data, but we found no public source of green labels for individual bonds on these venues, and Nordic ABM does not report to FIRDS at all. 
A larger Nordic sample would therefore need licensed data, such as Stamdata or LSEG, which would also provide issue dates and full bond histories for Nordic ABM. 

Our exposure measure is based on recent spot prices, although a firm that issues debt with a maturity of several years may pay closer attention to forward prices for later delivery. 
If forward prices stayed below spot prices during the crisis, issuers may have regarded the increase as temporary, and this would weaken any financing response to the measure used here. 
We were unable to obtain daily Nordic power futures for 2022 under terms that permit research use, so a direct test with forward prices remains an important extension. 

There are also limitations in coverage, timing and the assignment of exposure. 
The sample includes only the regulated market of Oslo B{\o}rs, so the Swedish and Danish firms in our data are issuers that chose to list in Oslo and do not represent their home markets. 
FIRDS reports the first trading day rather than the issue date, and the median gap of about 65 days for the ten bonds with a public issue date suggests that part of the price window may follow the decision to borrow. 
The zone of the headquarters is also an imperfect proxy for firms that operate in several zones, and restricting the sample to regional utilities leaves only six identifying issuers, which is too few for meaningful inference. 

Finally, the fixed effects and the placebo test do not rule out unobserved factors that vary by zone and quarter, so the estimates should be read as conditional associations and not as causal effects. 
The green label comes from an exchange classification, which we checked against issuer documents for five issuers without finding a wrongly labelled green bond. 
The check does not cover every bond, however, and two green bonds issued in 2026 were missing from the Euronext list. 

\section{Conclusion}

Green bonds increased from 15\% to 48\% of new green and conventional bond listings on Oslo B{\o}rs between 2019 and 2024. Their share remained around 36\% in 2022, even as electricity prices rose sharply and new listings declined by about one-third. Yet the within-issuer analysis provides no statistically significant evidence that higher local electricity prices changed the choice between green and conventional bonds once common quarterly developments were accounted for.

The findings are consistent with green bond issuance being shaped by market-wide developments and established financing practices rather than responding immediately to local energy costs, although the study cannot distinguish this interpretation from others, including investor demand. Green issuance is especially common among utilities, but the data do not establish why these firms choose the green label. The analysis also shows the value of retaining matured bonds when describing historical issuance trends, since currently listed instruments alone may give a distorted picture of earlier years.

The results need to be interpreted cautiously because only 21 issuers identify the main coefficient, and the confidence interval remains wide. Alternative price windows and tests among issuers with an existing green framework do not reveal a robust response, but adjustment beyond the observed period cannot be ruled out. Broader Nordic bond histories, together with issue dates, forward electricity prices, and investor-level data, would help establish whether green bond choice responds to energy price shocks under other market conditions.

\section*{Data availability}

Bond reference data are available from ESMA FIRDS, green bond classifications from Euronext, issuer information from GLEIF and the Br{\o}nn{\o}ysund Register Centre, electricity prices from the ENTSO-E Transparency Platform, policy rates from Norges Bank, and exchange rates from the ECB. The carbon price series is supplied by Ember using underlying Montel data, and permission for redistribution remains under review. Code for reconstructing the dataset from these sources will be linked when the repository is available.



\section*{Author contributions}
M.T.D.P.\ conceived the study, built the bond universe from ESMA FIRDS and the
matching of issuers to registries and bidding zones, designed and ran the
analysis, and wrote the paper. T.T.T., H.P.H.\ and D.T.N.\ provided the
electricity price data from the joint project on Nordic price formation and
commented on the design and the draft. All authors read and approved the
final version.


\section*{Acknowledgements}


\bibliographystyle{plainnat}
\bibliography{references}

\end{document}

%% file: tables/table1.tex
\begin{tabular}{lrrrrrrrr}
\toprule
 & Issuers & Bonds & Green & Conventional & Other ESG & Green share (count, \%) & Green share (EUR, \%) & Volume (EUR bn) \\
\midrule
\multicolumn{9}{l}{\textit{A. Listing year}} \\
2019 & 46 & 65 & 10 & 55 & 0 & 15.4 & 11.3 & 4.5 \\
2020 & 50 & 82 & 23 & 59 & 0 & 28.0 & 28.8 & 6.1 \\
2021 & 55 & 103 & 34 & 61 & 8 & 35.8 & 35.1 & 9.9 \\
2022 & 48 & 70 & 24 & 42 & 4 & 36.4 & 27.4 & 7.3 \\
2023 & 41 & 67 & 25 & 38 & 4 & 39.7 & 29.0 & 5.7 \\
2024 & 56 & 107 & 50 & 54 & 3 & 48.1 & 37.2 & 10.2 \\
2025 & 64 & 105 & 42 & 60 & 3 & 41.2 & 26.3 & 12.3 \\
\addlinespace
\multicolumn{9}{l}{\textit{B. Issuer type}} \\
Utility & 18 & 127 & 95 & 32 & 0 & 74.8 & 79.4 & 8.5 \\
Utility producers & 12 & 65 & 54 & 11 & 0 & 83.1 & 89.4 & 4.7 \\
Utility grid companies & 3 & 11 & 6 & 5 & 0 & 54.5 & 47.8 & 0.9 \\
Utility integrated firms & 3 & 51 & 35 & 16 & 0 & 68.6 & 73.2 & 2.8 \\
Real estate & 14 & 109 & 40 & 68 & 1 & 37.0 & 44.9 & 7.0 \\
Other & 155 & 363 & 73 & 269 & 21 & 21.3 & 14.9 & 40.3 \\
\addlinespace
\multicolumn{9}{l}{\textit{C. HQ country}} \\
Norway & 119 & 471 & 174 & 282 & 15 & 38.2 & 34.1 & 41.1 \\
Sweden & 9 & 39 & 27 & 10 & 2 & 73.0 & 67.7 & 2.3 \\
Denmark & 6 & 13 & 0 & 13 & 0 & 0.0 & 0.0 & 1.2 \\
Finland & 1 & 1 & 0 & 1 & 0 & 0.0 & 0.0 & 0.0 \\
Germany/Lux. & 5 & 6 & 1 & 5 & 0 & 16.7 & 9.9 & 1.0 \\
Other countries & 47 & 69 & 6 & 58 & 5 & 9.4 & 6.1 & 10.3 \\
\addlinespace
\multicolumn{9}{l}{\textit{Total}} \\
All & 187 & 599 & 208 & 369 & 22 & 36.0 & 29.2 & 55.9 \\
\bottomrule
\end{tabular}